\documentclass[11pt]{amsart}
\usepackage{amssymb,amsthm,latexsym}
\newcommand{\Z}{\mathbb{Z}}
\newcommand{\Q}{\mathbb{Q}}

\newcommand{\rationals}{\Q}
\newcommand{\integers}{\Z}

\newcommand{\set}[1]{\left\{#1\right\}}

\DeclareMathOperator{\Ideal}{Ideal}

\def\cocoa{{\hbox{\rm C\kern-.13em o\kern-.07em C\kern-.13em
o\kern-.15em A}}}

\newcommand{\design}{{\mathcal D}}
\newcommand{\fraction}{{\mathcal F}}
\newcommand{\idealof}[1]{\Ideal\left(#1\right)}

\title[Union of regular fractions]{2-level fractional factorial designs which are the union of non trivial regular designs}
\author[R. Fontana and G. Pistone]{Roberto Fontana and Giovanni Pistone}
\date{Presented by R. Fontana at the DAE 2007 Conference, The University of Memphis, November 2, 2007}
\address{DIMAT Politecnico di Torino}
\email{\{giovanni.pistone|roberto.fontana\}@polito.it}
\newtheorem{theorem}{Theorem}[section]
\newtheorem{corollary}{Corollary}[theorem]
\newtheorem{proposition}{Proposition}[section]
\newtheorem{definition}{Definition}[section]

\begin{document}\maketitle
\begin{abstract}
Every fraction is a union of points, which are trivial regular fractions. To characterize non trivial decomposition, we derive a condition for the inclusion of a regular fraction as follows. 
Let $F = \sum_\alpha b_\alpha X^\alpha$ be the indicator polynomial of a generic fraction, see Fontana et al, JSPI 2000, 149-172. Regular fractions are characterized by $R = \frac 1l \sum_{\alpha \in \mathcal L} e_\alpha X^\alpha$, where $\alpha \mapsto e_\alpha$ is an group homeomorphism from $\mathcal L \subset \mathbb Z_2^d$ into $\{-1,+1\}$. The regular $R$ is a subset of the fraction $F$ if $FR = R$, which in turn is equivalent to $\sum_t F(t)R(t) = \sum_t R(t)$. If $\mathcal H = \{\alpha_1 \dots \alpha_k\}$ is a generating set of $\mathcal L$, and $R = \frac1{2^k}\left(1 + e_1X^{\alpha_1}\right) \cdots \left(1 + e_kX^{\alpha_k}\right)$, $e_j = \pm 1$, $j=1 \dots k$,  the inclusion condition in term of the $b_\alpha$'s is
\begin{equation}b_0 + e_1 b_{\alpha_1} + \cdots + e_1 \cdots e_k b_{\alpha_1 + \cdots + \alpha_k} = 1. \tag{*}\end{equation}
The last part of the paper will discuss some examples to investigate the practical applicability of the previous condition (*). 

\emph{This paper is an offspring of the Alcotra 158 EU research contract on the planning of sequential designs for sample surveys in tourism statistics.}
\end{abstract}
\section{Introduction}
We consider 2-level fractional designs with $m$ factors, where the levels of each factor are coded $-1,+1$. The full factorial design is $\mathcal D = \set{-1,+1}^m$ and a fraction of the full design is a subset $\mathcal F \subset \mathcal D$. According to the algebraic description of designs, as it is discussed in \cite{pistone|riccomagno|wynn:2001}, \cite{pistone|riccomagno|rogantin:07henry}, the \emph{fraction ideal} $\idealof\fraction$, also called design ideal, is the set of all polynomials with real coefficients that are zero on all points of the fraction. Two polynomials $f$ and $g$ are aliased by $\fraction$ if and only if $f - g \in \idealof\fraction$ and the quotient space defined in such a way is the vector space of real responses on $\fraction$. The fraction ideal is generated by a finite number of its elements. This finite set of polynomials is called a \emph{basis} of the ideal. bases are not uniquely determined, unless very special conditions are met. A \emph{Gr\"obner basis} of the fraction ideal can be defined after the assignment of a total order on monomials called \emph{monomial order}. If a monomial order is given, it is possible to identify the \emph{leading monomial} of each polynomial. As far as applications to statistics are concerned, a Gr\"obner basis is characterized by the following property: the set of all monomials that are are not divided by any of the leading term of the polynomials in the basis form a linear basis of the quotient vector space. A general reference to the relevant computational commutative algebra topics is \cite{cox|little|oshea:1997}.

The ring of polynomials in $m$ indeterminates $x_1 \dots x_m$ and rational coefficient is denoted by $R = \rationals\left[x_1 \dots x_m\right]$. The design ideal $\idealof\design$ has a unique `minimal' basis $x_1^2-1,\dots,x_m^2-1$, which happens to be a Gr\"obner basis. The polynomials that are added to this basis to generate the ideal of a fraction are called \emph{generating equations}. An ideal with a basis of binomials with coefficients $\pm 1$ is called \emph{binomial ideal}. Indicator polynomials polynomials of a fraction were introduced in \cite{fontana|pistone|rogantin:2000}, see also \cite{pistone|rogantin:2007-JSPI}. An indicator polynomial has the form
\begin{equation}
  \label{eq:indicator}
  F = \sum_{\alpha} b_\alpha x^\alpha, \quad \alpha = (\alpha_1,\dots, \alpha_m) \in \set{0,1}, \quad x^\alpha = x_1^{\alpha_1} \cdots x_d^{\alpha_d}
\end{equation}
and it satisfies the conditions $F(a)=1$ if $a\in \fraction$, $F(a)=0$ otherwise. If necessary, we distinguish between the indeterminate $x_j$, the value $a_j$ and the mapping $X_j(a) = a_j$. How to move between the ideal representation and the indicator function representation, is discussed in \cite{notari|riccomagno|rogantin:2007}.

The definition and characterization, from the algebraic point of view, of regular fractional factorial designs (briefly regular designs) is discussed in \cite{fontana|pistone|rogantin:2000}, see also \cite{pistone|rogantin:2007-JSPI}. In particular, the last paper referred to considers mixed factorial design, but this case is outside the scope of the present paper. Orthogonal arrays as are defined in \cite{hedayat|sloane|stufken:1999} can be characterized in the previous algebraic framework, see \cite{pistone|rogantin:2007-JSPI} and \cite{carlini|pistone:2007}, as follows. A fraction $\fraction$ with indicator polynomial $F$ is orthogonal with strength $s$ if $b_\alpha=0$ if $1 \le |\alpha| \le s$,  $|\alpha| = \sum_j \alpha_j$. The notion of indicator polynomial can be accommodated to cases with replicated design points by allowing integer values other than 0 and 1 to $F$, see \cite{ye:2003}. In such a case, we prefer to call $F$ a \emph{counting polynomial} of the fraction. A systematic algebraic search of orthogonal arrays with replications is discussed in \cite{carlini|pistone:2007}. For sake of easy reference in Section \ref{sec:pb} below, we quote a couple of specific result about orthogonal arrays. In fact, considering $m=5$ factors and strength $s=2$, it is shown in \cite[Table 5.2]{carlini|pistone:2007} that there are 192 OA's with 12 points and no replications, and there are 32 OA's with 12 points, one of them replicated.   

This paper is organized as follows. In Section \ref{regfrac} the algebraic theory is reviewed and in Section \ref{sec:union} it is applied to the problem of finding fractions that are union of regular fractions. In Section \ref{sec:pb} the important case of Plackett-Burman designs is considered.
\section{Regular fractions}\label{regfrac}
According to the definitions in \cite{fontana|pistone|rogantin:2000} and \cite{pistone|rogantin:2007-dimat2} a regular fraction is defined as follows. Let $\mathcal L$ be a subset of $L = \integers_2^m$, which is an additive group. Let $\Omega_2$ be the multiplicative group $\set{-1,+1}$
\begin{definition} \label{def:reg}
Let $e$ be a map from $\mathcal L$ to $\Omega_2$.
A non-empty fraction $\fraction$ is \emph{regular} if
\begin{enumerate}
 \item $\mathcal L \subset L$ s a sub-group;
 \item the equations
\begin{equation*} %\label{eq:def-eq}
X^\alpha = e(\alpha) \quad , \qquad  \alpha \in \mathcal L
\end{equation*}
define the fraction $\fraction$, i.e. are a set of \emph{generating equations}.
\end{enumerate}
In such a case, $e$ is a group homeomorphism.
\end{definition}

Other known definitions are shown to be equivalent to this one by the following proposition.
\begin{theorem}\label{th:portmaneaux}
Let $\fraction$ be a fraction.  The following statements are equivalent:
\begin{enumerate}
 \item \label{it:reg} The fraction $\fraction$ is regular according to  definition \ref{def:reg}.
  \item \label{it:reg-F} The indicator function of the fraction has the form
\begin{equation*}
F(\zeta)=\frac 1 l \sum_{\alpha \in \mathcal L} e(\alpha)\ X^\alpha(\zeta), \qquad \zeta \in \design.
\end{equation*}
where $\mathcal L$ is a given subset of $L$ and $e: \mathcal L \to
\Omega_2$ is a given mapping.
 \item \label{it:reg-con} For each $\alpha, \beta \in  L$ the interactions represented on $\fraction$ by the
 terms $X^\alpha$  and $X^\beta$ are either orthogonal or totally aliased.
\item \label{it:binomial}The $\idealof{\fraction}$ is binomial.
\item \label{it:group} $\fraction$ is either a subgroup or a lateral of a subgroup of the multiplicative group $\design$
\end{enumerate}
\end{theorem}
\begin{proof}
 Most of the equivalences are either well known or proved in the cited literature. We prove the equivalence of \eqref{it:binomial} The ideal of a regular design is generated by the basis of the full design and by generating polynomials of the form $X^\alpha-e_\alpha$, where $e_\alpha = \pm 1$; all these polynomials are binomials. Viceversa, if the variety of a binomial ideal is a fraction of $\design$, then all the polynomials $x_i^2-1$ are contained in its ideal, and every other binomial in the basis, say $x^\alpha-ex^\beta$, $e=\pm 1$, is equivalent to the generating polynomial $x^{\alpha+\beta}+e$.
\end{proof}

We will show some examples of application of such theorem below. We first will prove two propositions that characterize the simple cases of \emph{1-point} and \emph{2-points} regular fractions.

\begin{proposition}
Every 1-point fraction is regular
\end{proposition}

\begin{proof}
We can prove the statement using design ideals.  A single generic point is $a = (a_1,\dots, a_m) \in \design$. A binomial basis is $\set{x_i-a_i, i=1,\dots,m}$ and, therefore, $\fraction \equiv \left\{a \right\}$ is regular. 

Equivalently we can use indicator functions. Indeed the indicator function of a single point $a$ is $F_a = \frac 1{2^m} (1+a_1x_1) \cdot \cdots \cdot (1+a_mx_m)$ and $F_a$ meets the requirements for being an indicator function of a regular design.
\end{proof}

% magari da recuperare pi avanti
%Remark
%Given $\mathcal F \subseteq \mathcal D$ and a point $p(p_1 \cdots p_m) \in \mathcal F$ and indicating with $F$ the indicator function of $\mathcal F$ we have
%\begin{equation}
%F(p)=1 \Leftrightarrow b_0 + b_1 p_1 + \cdots b_m p_m + b_{12} p_1 p_2 + \cdots + b_{12 \cdots m} p_1 \cdots p_m \nonumber
%\end{equation} 

The following result looks less trivial.

\begin{proposition}
Every 2-points fraction is regular.
\end{proposition}

\begin{proof}
Let $ \underline{1}=(1,\cdots,1)$ be the null element of $\design$. We observe that every subset $\fraction$ of $\design$ made up of two elements, say $a$ and $b$ with $a\neq b$ is a subgroup or a coset of a subgroup. Indeed if $a=\underline{1}$ or $b=\underline{1}$ then $\fraction$ is a subgroup. If $a \neq \underline{1}$ and $b \neq \underline{1}$ then $\fraction$ is the coset $a H$ where $H$ is the subgroup $ \set{\underline{1},a^{-1}b} $.
\end{proof}

\begin{subsection}{Remark}
We can also prove the result comparing  the number of 2-points subsets with the number of subgroups of order 2. The number of 2-points fractions of $\mathcal D$ is

\begin{equation*}
\left( \begin{array}{cc} 2^m \\ 2 \end{array} \right) = \frac{2^m \cdot (2^m -1)}{2} = 2^{m-1} \cdot (2^m - 1)  
\end{equation*}

On the other side, every regular fraction is a subgroup of $\mathcal D$ or a coset of a subgroup of $\mathcal D$ (\cite{fontana|pistone|rogantin:2000}). In particular 
 the number of regular fractions of size 2 is equivalent to the number of subgroups of order $2$ multiplied by the number of cosets of a subgroup, that is $2^{m-1}$.
 
The number of subgroups of order equal to $2$ is  $2^m - 1$. Indeed every set $ \set{\underline{1},p} $ with $ \underline{1}=(1,\cdots,1)$ and $p \in \mathcal D, p \neq \underline{1}$ is a subgroup of order equal to 2.

It follows that the number of regular fractions of size 2 will be equal to

\begin{equation*}
2^{m-1} \cdot (2^m - 1)
\end{equation*}

that is the number of 2-points fraction.

If we consider $2^k$-points fractions ($k \geq 2$) a similar argument is not valid as will be clear in the next sections.

It also follows that every 3-points fraction can be considered as the union of a 1-point fraction and a 2-points fraction.

\end{subsection}

\section{Union of regular designs} \label{sec:union}
In this section we consider the union of regular designs. To simplify formul{\ae} we will introduce the following notation:
\[
X^{\alpha} \equiv X_1^{\alpha_1} \cdot \dots \cdot X_m^{\alpha_m} = X_{\bar{\alpha}}
\]
where $\bar{\alpha}$ is the set for which $\alpha_i \neq 0$, $\left \{ i \in \left \{ 1, \dots , m \right\} : \alpha_i \neq 0 \right\}$. We will also write $\alpha$ in place of $\bar{\alpha}$ with a small abuse of notation. As an example let's consider $m=4$ and $\alpha=(0,1,1,0)$.
It follows that $X^{\alpha}=X_2 X_3$ will be written as $X_{23}$.

Let $\mathcal F_1 $ and $\mathcal F_2$ two regular fractions, both included in $\mathcal D$. The indicator functions of $\mathcal F_1$ and $\mathcal F_2$, say $F_1$ and $F_2$ respectively, allow to easily determine the indicator function of the union of $\mathcal F_1$ and $\mathcal F_2$, $\mathcal F = \mathcal F_1 \cup \mathcal F_2$ as
\begin{equation*} F = F_1 + F_2 - F_1 \times F_2 \end{equation*}

In general, the union of two (disjoint) regular fractions is not a regular fraction. As an example let's consider $m=2$ factors, $\mathcal D = \set{-1,+1}  \times \set{-1,+1}$ and $\mathcal F_1 = \set{(-1,-1)}$ and  $\mathcal F_2 = \set{(-1,+1), (+1,-1))}$. 
Both $\mathcal F_1$ and $\mathcal F_2$ are regular fractions, according to the propositions of the previous sections. Indeed their indicator functions meet the requirements for regular fractions: $F_1 = \frac{1}{4} ( 1- X_1) \cdot (1-X_2)$ and $F_2 = \frac{1}{2} ( 1- X_1 \cdot X_2)$. However, the union 
 $\mathcal F = \set{(-1,-1), (-1,+1), (+1,-1) )}$, is not a regular fraction, because  its indicator function is 
 \begin{equation*}
   F = \frac{3}{4} - \frac{1}{4} X_1 - \frac{1}{4} X_2 - \frac{1}{4} X_1 \cdot X_2.
 \end{equation*}

The same conclusion can be obtained considering \textit{design ideals} related to fractional designs. Given $\mathcal F_1 \subset \mathcal D$, $\mathcal F_2 \subset \mathcal D$ and $\mathcal F = \mathcal F_1 \cup \mathcal F_2$ the associated ideals will be $\idealof {\mathcal F_1} $, $\idealof { \mathcal F_2 }$ and $\idealof {  \mathcal F} $. In general, the fact that $\idealof { \mathcal F_1}$ and $\idealof { \mathcal F_2}$ are binomial ideals by Theorem \ref{1{th:portmaneaux}} doesn't imply that $\idealof { \mathcal F)} $ is a binomial ideal . Indeed, for the previous example, the Gr\"obner bases $B_1$, $B_2$ and $B$ of  $\idealof { \mathcal F_1} $, $\idealof { \mathcal F_2} $ and $\idealof { \mathcal F} $   respectively, are:
\begin{equation*}
\begin{aligned}
B_1 &= \set { X_1 + 1, X_2 + 1} \\ 
B_2 &= \set { X_2^2 - 1, X_1 + X_2} \\ 
B   &= \set { -1/4 X_1 X_2 - 1/4 X_1 - 1/4 X_2 - 1/4, X_2^2 - 1, X_1^2 - 1}
\end{aligned}
\end{equation*}
It results that $\idealof { \mathcal F_1} $ and $\idealof { \mathcal F_2} $ are binomial ideals while $\idealof { \mathcal F} $ is not.

\subsection{Remark}
More generally, let's consider two disjoint regular fractions, namely $aG$ and $bH$, where $G$ and $H$ are subgroups of $\design$ and $a \notin G$ and $b \notin H$. Let's take $ag$ and $bh$. In order to have $(ag)(bh) \in aG$ we should have $bgh \in G$ or, equivalently, $bh \in G$.     
\section{Decomposing a fraction into regular fractions}
In this part of the work we would like to explore the inverse path, i.e. to analyze the decomposition of a given $\mathcal F \subset \mathcal D$ into the union of disjoint regular fractions. We will indicate with $\mathcal R$ the generic regular fraction.

Let's indicate with $F$ and $R$ the indicator functions of $\mathcal F \subset \mathcal D$ and $\mathcal R \subset \mathcal D$ respectively. Under which condition $\mathcal R$ will be a subset of  $\mathcal F$?

\begin{theorem}\label{th:regsumbi}
Let $F$ be the indicator function of a generic fractional design $\mathcal F \subset \mathcal D$, $F = \sum_\alpha b_\alpha X^\alpha$. 
Let $R$ the indicator function of a regular fractional design $\mathcal R \subset \mathcal D$, $R = \frac 1l \sum_{\alpha \in \mathcal L} e_\alpha X^\alpha = \frac1{2^k}\left(1 + e_1X^{\alpha_1}\right) \cdots \left(1 + e_kX^{\alpha_k}\right)$ .
The following statement holds:
\begin{equation} \mathcal R \subseteq \mathcal F \Leftrightarrow  b_0 + e_1 b_{\alpha_1} + \cdots + e_1 \cdots e_k b_{\alpha_1 + \cdots + \alpha_k} = 1 \nonumber
\end{equation}
\end{theorem}

\begin{proof}
For $\mathcal R$ to be a subset of $\mathcal F$ it must happen that the number of points of  $\mathcal R$ must be equal to the number of points of $\mathcal R \cap \mathcal F $. In terms of indicator functions the equality $\mathcal R = \mathcal R \cap \mathcal F $ becomes  $\sum_t F(t)R(t) = \sum_t R(t)$ being $ t \in \mathcal D$.
We have

\begin{multline*}
FR = \\ (\sum_\alpha b_\alpha X^\alpha) \cdot \frac1{2^k}\left(1 + e_1X^{\alpha_1}\right) \cdots \left(1 + e_kX^{\alpha_k}\right) = \\ 
\frac1{2^k} \sum_\alpha b_\alpha X^\alpha + \frac1{2^k} \sum_\alpha b_\alpha X^\alpha e_1 X^{\alpha_1} \cdots + \frac1{2^k} \sum_\alpha b_\alpha X^\alpha e_1 \cdots e_k  X^{\alpha_1 + \cdots +\alpha_k }
\end{multline*}

It follows that

\begin{equation}
\sum_t F(t)R(t) = \frac1{2^k} 2^m b_0 +  \frac1{2^k} 2^m e_1 b_{\alpha_1} + \cdots \frac1{2^k} 2^m e_1 \cdots e_k b_{\alpha_1 + \cdots +\alpha_k  } \nonumber
\end{equation}

On the other hand
\begin{equation}
\sum_t R(t) = \frac1{2^k} 2^m \nonumber
\end{equation}

It follows 

\begin{equation}
b_0 + e_1 b_{\alpha_1} + \cdots + e_1 \cdots e_k b_{\alpha_1 + \cdots + \alpha_k} = 1 \nonumber
\end{equation} \end{proof} 

\begin{corollary}
A necessary, but not sufficient, condition for a regular fraction $\mathcal R$ to be contained in $\design$ is 
\[
b_0 + \left| b_{\alpha_1}\right| + \cdots + \left| b_{\alpha_1 + \cdots + \alpha_k} \right| \geq 1 \nonumber
\] 
\end{corollary}

\subsection{A small example}
Let's consider the 3-points fraction $\mathcal F \subset \mathcal D = \set{-1,+1}  \times \set{-1,+1}$ that we have introduced in the previous section:
\begin{equation*}
\mathcal F= \set{(-1,-1), (-1,+1), (+1,-1) )}
\end{equation*}
The indicator function $F$ of $\mathcal F$ is $F = \frac{3}{4} - \frac{1}{4} X_1 - \frac{1}{4} X_2 - \frac{1}{4} X_1 \cdot X_2$, that is $b_0=\frac{3}{4}, b_1=- \frac{1}{4}, b_2=- \frac{1}{4}, b_{12}= - \frac{1}{4}$. 
It follows

\begin{eqnarray*}
b_0 - b_1 = 1 \\
b_0 - b_2 = 1 \\
b_0 - b_{12} = 1 \\
b_0 - b_1 - b_2 + b_{12}= 1 \\ 
b_0 - b_1 + b_2 - b_{12}= 1 \\ 
b_0 + b_1 - b_2 - b_{12}= 1 
\end{eqnarray*}

From each relation, using theorem \ref{th:regsumbi}, we can obtain the indicator functions of the regular fractions that are contained into $\mathcal F$.
These are 
\begin{eqnarray*}
F_1 = \frac{1}{2} ( 1- X_1) \\
F_2 = \frac{1}{2} ( 1- X_2) \\
F_3 = \frac{1}{2} ( 1- X_1 \cdot X_2) \\
F_4 = \frac{1}{4} ( 1- X_1) \cdot (1-X_2) \\
F_5 = \frac{1}{4} ( 1- X_1) \cdot (1 + X_2) \\
F_6 = \frac{1}{4} ( 1+ X_1) \cdot (1-X_2) 
\end{eqnarray*}
 
Therefore the corresponding regular fractions are, respectively

\begin{eqnarray*}
\mathcal F_1 = \set{(-1,-1), (-1,+1))} \\
\mathcal F_2 = \set{(-1,-1), (+1,-1))} \\
\mathcal F_3 = \set{(-1,+1), (+1,-1))} \\
\mathcal F_4 = \set{(-1,-1)} \\
\mathcal F_5 = \set{(-1,+1)} \\
\mathcal F_6 = \set{(+1,-1)} 
\end{eqnarray*}

\section{Plackett-Burman designs} \label{sec:pb}
Another example can be obtained considering the well-known ``Plackett-Burman'' designs \cite{plackett|burman:46}. In particular the Plackett-Burman design for 11 variables and 12 runs is built according the following procedure:

\begin{enumerate}
\item the first row, namely the key, is given: $+ + - + + + - - - + -$
\item the second row up to the eleventh row are built shifting the key of one position each time
\item the 12th row is set equal to $- - - - - - - - - - -$
\end{enumerate}
The Plackett-Burman design for eleven parameters becomes

\[
\begin{array}{crrrrrrrrrrr}
N & A & B & C & D & E & F & G & H & I & J & K \\
1 & + & + & - & + & + & + & - & - & - & + & - \\
2 & - & + & + & - & + & + & + & - & - & - & + \\
3 & + & - & + & + & - & + & + & + & - & - & - \\
4 & - & + & - & + & + & - & + & + & + & - & - \\
5 & - & - & + & - & + & + & - & + & + & + & - \\
6 & - & - & - & + & - & + & + & - & + & + & + \\
7 & + & - & - & - & + & - & + & + & - & + & + \\
8 & + & + & - & - & - & + & - & + & + & - & + \\
9 & + & + & + & - & - & - & + & - & + & + & - \\
10 & - & + & + & + & - & - & - & + & - & + & + \\
11 & + & - & + & + & + & - & - & - & + & - & + \\
12 & - & - & - & - & - & - & - & - & - & - & -
\end{array}
\]

We consider the case with $m=5$ factors and, from the ``Plackett-Burman'' for 11 factors we randomly select the following $\mathcal F$, corresponding to columns A,B,F,H and I of the original design. 

% da spostare 
% There are 70 different ``Plackett-Burman'' designs for 5 factors and with 12 runs, as we determined exploring, using a routine written in SAS IML, all the $\left( \begin{array}{c} 11 \\ 5 \end{array} \right) = 462$ choice of five columns out of eleven. 

The plus sign '$+$' has been coded with '$1$' and the minus sign '$-$' with '$-1$'. 

\[
\begin{array}{cc}
\mathcal F = &
\begin{array}{crrrrr}
N & X_1 & X_2 & X_3 & X_4 & X_5 \\
1 & 1 & 1 & 1 & 1 & 1 \\
2 & 1 & 1 & -1 & -1 & 1 \\
3 & 1 & -1 & -1 & -1 & 1 \\ 
4 & -1 & 1 & -1 & 1 & 1 \\
5 & -1 & -1 & 1 & 1 & 1 \\
6 & -1 & -1 & 1 & -1 & 1 \\
7 & 1 & 1 & 1 & -1 & -1 \\
8 & 1 & -1 & 1 & 1 & -1 \\
9 & 1 & -1 & -1 & 1 & -1 \\
10 & -1 & 1 & 1 & -1 & -1 \\
11 & -1 & 1 & -1 & 1 & -1 \\
12 & -1 & -1 & -1 & -1 & -1
\end{array}
\end{array}
\]

The indicator function $F$ of $\fraction$ is

{
\begin{multline*}
\frac{ 3 }{ 8 } + \frac{ 1 }{ 8 } X_{ 3 4 5 } + \frac{ 1 }{ 8 } X_{ 2 4 5 } - \frac{ 1 }{ 8 } X_{ 2 3 5 } - \frac{ 1 }{ 8 } X_{ 2 3 4 } + \\ \frac{ 1 }{ 8 } X_{ 2 3 4 5 } - \frac{ 1 }{ 8 } X_{ 1 4 5 } - \frac{ 1 }{ 8 } X_{ 1 3 5 } + \frac{ 1 }{ 8 } X_{ 1 3 4 } + \frac{ 1 }{ 8 }  X_{ 1 3 4 5 } + \frac{ 1 }{ 8 } X_{ 1 2 5 } + \\
- \frac{ 1 }{ 8 } X_{ 1 2 4 } + \frac{ 1 }{ 8 } X_{ 1 2 4 5 } + \frac{ 1 }{ 8 } X_{ 1 2 3 } + \frac{ 1 }{ 8 } X_{ 1 2 3 5 } + \frac{ 1 }{ 8 } X_{ 1 2 3 4 }
\end{multline*}
}

It follows that $\mathcal F$ is not regular.

Now we start to search for regular fractions that are contained in $\mathcal F$. 

Of course the first constraint concerns the size of the regular fraction. It must be less or equal to 12, the number of points of $\mathcal F$. Being $\mathcal R$ a regular fraction, it follows that the size of $\mathcal R$ could be $2^0=1$ or $2^1=2$ or $2^2=4$ or $2^3=8$.

We already know, from the propositions of section \ref{regfrac} that

\begin{itemize} 
\item all the 12 points that constitute $\mathcal R$ are 1-point regular fraction;
\item all the $\left( \begin{array}{cc} 12 \\ 2 \end{array} \right) = 66 $ are 2-points regular fraction. 
\end{itemize}

Let's study 4-points and 8-points subsets of $\mathcal F$.

The corollary of theorem Th. \ref{th:regsumbi} allows us to immediately exclude 8-points regular fractions. Indeed the following condition should be true for a proper choice of $e_1, e_2$ and $\alpha_1, \alpha_2$   

\begin{equation*}b_0 + e_1 b_{\alpha_1} + e_2 b_{\alpha_2} +  e_1 e_2 b_{\alpha_1 + \alpha_2} = 1 \end{equation*}

But $b_0 = \frac{ 3 }{ 8 }$ and the absolute value of $b_i$ is $\frac{ 1 }{ 8 }, \forall i$ and so it is not possible that the left  side of the previous equation sums up to 1. No 8-points regular fraction is contained into $\fraction$.

Finally we investigate 4-points regular fractions.

For a proper choice of $e_1, e_2, e_3$ and $\alpha_1, \alpha_2, \alpha_3$ the following relation should hold

\begin{equation*}b_0 + e_1 b_{\alpha_1} + e_2 b_{\alpha_2} + e_3 b_{\alpha_3} + e_1 e_2 b_{\alpha_1 + \alpha_2} 
 + e_1 e_3 b_{\alpha_1 + \alpha_3} + e_2 e_3 b_{\alpha_2 + \alpha_3} + e_1 e_2 e_3 b_{\alpha_1 + \alpha_2 + \alpha_3}= 1 \end{equation*}  

A subgroup of order eight will be $\left\{ \underline {1}, a, b, ab, c, ac, bc, abc \right\}$ with $a \neq \underline {1}$, $b \neq \underline {1}$, $c \neq \underline {1}$  and $a \neq b$, $a \neq c$ and $b \neq c$. We can choose $a$, $b$ and $c$ in $\left( \begin{array}{c} 31 \\ 2 \end{array} \right) \cdot (31-3)$ different ways. The number of different subgroups is obtained dividing this number by  $\left( \begin{array}{c} 7 \\ 2 \end{array} \right) \cdot 4$. We get 155 different subgroups. 
 
% Direi che non serve più
%A subgroup of order four will be $\left\{ \underline {1}, a, b, ab \right\}$ with $a \neq \underline {1}$, $b \neq \underline {1}$ and $a \neq b$. We can choose $a$ and $b$ in $\left( \begin{array}{c} 31 \\ 2 \end{array} \right)$. The number of different subgroups is obtained dividing this number by  $\left( \begin{array}{c} 3 \\ 2 \end{array} \right)$. We get 155. 

Every subgroup of order 8, $\mathcal S_i^{(8)}=< \alpha_{1i}, \alpha_{2i}, \alpha_{3i} >, i=1, \dots , 155$ defines 8 regular fractions of size 4 (the subgroup orthogonal to $\mathcal S_i^{(8)}$ and its cosets). To find the regular fractions embedded into $\mathcal F$ we must solve the following systems of equations $(i=1,\dots,155)$ 

{\tiny 
\[
\left\{
\begin{array}{l}
e_1^2 - 1 = 0 \\
e_2^2 - 1 = 0 \\
e_3^2 - 1 = 0 \\
b_0 + e_1 b_{\alpha_{1i}} + e_2 b_{\alpha_{2i}} + e_3 b_{\alpha_{3i}} + e_1 e_2 b_{\alpha_{1i} + \alpha_{2i}} 
 + e_1 e_3 b_{\alpha_{1i} + \alpha_{3i}} + e_2 e_3 b_{\alpha_{2i} + \alpha_{3i}} + e_1 e_2 e_3 b_{\alpha_{1i} + \alpha_{2i} + \alpha_{3i}} - 1=0
\end{array}
\right.
\]
}
To do it we generate the 155 subgroups of $\mathcal D$ of order eight (for example using the package GAP \cite{GAP4}). 
As an example let's consider $\mathcal S_1=< \left\{1\right\} , \left\{2\right\}, \left\{3\right\} >$. Being $b_0=\frac{ 3 }{ 8 }$, $b_1=b_2=b_3=b_{12}=b_{13}=b_{23}=0$ and $b_{123}= \frac{ 1 }{ 8 }$ the corresponding system of equation is 
\[
\left\{
\begin{array}{l}
e_1^2 - 1 = 0 \\
e_2^2 - 1 = 0 \\
e_3^2 - 1 = 0 \\
\frac{ 3 }{ 8 } + \frac{ 1 }{ 8 } e_1 e_2 e_3 -1=0 
\end{array}
\right.
\]

The system doesn't have any solution.

Let's now consider $\mathcal S_2=< \left\{4\right\} ,\left\{12\right\}, \left\{135\right\} >$. Being $b_0=\frac{ 3 }{ 8 }$, $b_4=b_12=0$, 
$b_{135}=b_{124}=b_{235}= - \frac{ 1 }{ 8 }$ and $b_{1345}=b_{2345}= \frac{ 1 }{ 8 }$ the corresponding system of equation is 
\[
\left\{
\begin{array}{l}
e_1^2 - 1 = 0 \\
e_2^2 - 1 = 0 \\
e_3^2 - 1 = 0 \\
\frac{ 3 }{ 8 } - \frac{ 1 }{ 8 } e_3 - \frac{ 1 }{ 8 } e_1 e_2 + \frac{ 1 }{ 8 } e_1 e_3 - \frac{ 1 }{ 8 } e_2 e_3 + \frac{ 1 }{ 8 } e_1 e_2 e_3-1=0 
\end{array}
\right.
\]

The system has the following solution $e_1=-1, e_2=1, e_3=-1$ that defines the following indicator function $F^{ (1) }$
\[
\frac{ 1 }{ 8 } (1-X_4)(1+X_{12})(1-X_{135})
\]

The corresponding set of points $\fraction^{ (1) }$ is

\[
\begin{array}{crrrrr}
N & X_1 & X_2 & X_3 & X_4 & X_5 \\
2 & 1 & 1 & -1 & -1 & 1 \\
6 & -1 & -1 & 1 & -1 & 1 \\
7 & 1 & 1 & 1 & -1 & -1 \\
12 & -1 & -1 & -1 & -1 & -1
\end{array}
\]

To proceed into the decomposition of $\fraction$ we remove the points of $\fraction^{ (1) }$. The indicator function of the new set will be 
$F - F^{ (1) }$:
{
\begin{multline*}
\frac{ 1 }{ 4 } + \frac{ 1 }{ 8 } X_{4 } - \frac{ 1 }{ 8 } X_{ 1 2 }  + \frac{ 1 }{ 8 } X_{ 3 4 5 } + \frac{ 1 }{ 8 } X_{ 2 4 5 }  - \frac{ 1 }{ 8 } X_{ 2 3 4 } + \\ 
- \frac{ 1 }{ 8 } X_{ 1 4 5 }  + \frac{ 1 }{ 8 } X_{ 1 3 4 }  + \frac{ 1 }{ 8 } X_{ 1 2 5 }  + \frac{ 1 }{ 8 } X_{ 1 2 4 5 } + \frac{ 1 }{ 8 } X_{ 1 2 3 } + \\
 + \frac{ 1 }{ 8 } X_{ 1 2 3 5 } + \frac{ 1 }{ 8 } X_{ 1 2 3 4 }
\end{multline*}
}

We now search for the regular fractions contained into $\fraction - \fraction^{ (1) }$. A regular fraction $\mathcal R$ to be contained into $\fraction - \fraction^{ (1) }$ must be contained into $\fraction$. We can therefore limit our search to the solution that we have identified in the first part. Let's now consider $\mathcal S_3=< \left\{12\right\} ,\left\{35\right\}, \left\{245\right\} >$.

Being $b_0^{ (1) }=\frac{ 1 }{ 4 }$, $b_{35}^{ (1) }=0$ $b_{245}^{ (1) }=b_{134}^{ (1) }=b_{1235}^{ (1) }=\frac{ 1 }{ 8 }$ and $b_{234}^{ (1) }=b_{145}^{ (1) }=b_{12}^{ (1) }= - \frac{ 1 }{ 8 }$ the corresponding system of equation is 
\[
\left\{
\begin{array}{l}
e_1^2 - 1 = 0 \\
e_2^2 - 1 = 0 \\
e_3^2 - 1 = 0 \\
\frac{ 1 }{ 4 } - \frac{ 1 }{ 8 } e_1 + \frac{ 1 }{ 8 } e_3 + \frac{ 1 }{ 8 } e_1 e_2 - \frac{ 1 }{ 8 } e_1 e_3 - \frac{ 1 }{ 8 } e_2 e_3
+ \frac{ 1 }{ 8 } e_1 e_2 e_3 -1=0 
\end{array}
\right.
\]

The system has the following solution $e_1=-1, e_2=-1, e_3=1$ that defines the following indicator function $F^{ (2) }$
\[
\frac{ 1 }{ 8 } (1-X_{12})(1-X_{35})(1+X_{245})
\]

The corresponding set of points $\fraction^{ (2) }$ is

\[
\begin{array}{crrrrr}
N & X_1 & X_2 & X_3 & X_4 & X_5 \\
3 & 1 & -1 & -1 & -1 & 1 \\ 
4 & -1 & 1 & -1 & 1 & 1 \\
8 & 1 & -1 & 1 & 1 & -1 \\
10 & -1 & 1 & 1 & -1 & -1 \\
\end{array}
\]

If we remove this set of points from $\fraction - \fraction_1$ we get the following indicator function 
$F^{ (3)}=F - F^{ (1) } - F^{ (2) }$:
{
\begin{equation*}
\frac{ 1 }{ 8 } + \frac{ 1 }{ 8 } X_{4 } + \frac{ 1 }{ 8 } X_{ 3 5 } + \frac{ 1 }{ 8 } X_{ 3 4 5 } 
  + \frac{ 1 }{ 8 } X_{ 1 2 5 }  + \frac{ 1 }{ 8 } X_{ 1 2 4 5 } + \frac{ 1 }{ 8 } X_{ 1 2 3 } 
 + \frac{ 1 }{ 8 } X_{ 1 2 3 4 }
\end{equation*}
}
or, equivalently,
\begin{equation*}
\frac{ 1 }{ 8 } (1 + X_{4 }) (1+ X_{ 3 5 }) ( 1+ X_{ 1 2 5 } )
\end{equation*}

and the corresponding set of points $\fraction^{ (3) }$

\[
\begin{array}{crrrrr}
N & X_1 & X_2 & X_3 & X_4 & X_5 \\
1 & 1 & 1 & 1 & 1 & 1 \\
5 & -1 & -1 & 1 & 1 & 1 \\
9 & 1 & -1 & -1 & 1 & -1 \\
11 & -1 & 1 & -1 & 1 & -1 \\
\end{array}
\] 

$F^{ (3)}$ meets the requirements to be an indicator function of a regular design. We have therefore decomposed $\fraction$ into three regular designs, $\fraction = \fraction_1 \cup \fraction_2 \cup \fraction_3$. 

\subsection{Decomposition of the given `Plackett-Burman' design into \emph{all} the unions of 4-points regular designs }
In this part we find all the possible decompositions of the given ``Plackett-Burman'' design. As described in the previous section, we consider all the 155 subgroups of order 8, $\mathcal S_i^{(8)}=< \alpha_{1i}, \alpha_{2i}, \alpha_{3i} >, i=1, \dots , 155$ and we search for the solution of the following systems of equations 

{\tiny 
\[
\left\{
\begin{array}{l}
e_1^2 - 1 = 0 \\
e_2^2 - 1 = 0 \\
e_3^2 - 1 = 0 \\
b_0 + e_1 b_{\alpha_{1i}} + e_2 b_{\alpha_{2i}} + e_3 b_{\alpha_{3i}} + e_1 e_2 b_{\alpha_{1i} + \alpha_{2i}} 
 + e_1 e_3 b_{\alpha_{1i} + \alpha_{3i}} + e_2 e_3 b_{\alpha_{2i} + \alpha_{3i}} + e_1 e_2 e_3 b_{\alpha_{1i} + \alpha_{2i} + \alpha_{3i}} - 1=0
\end{array}
\right.
\]
}

15 of these 155 systems of equations have a non-empty set of solutions. Each of these non-empty sets define an indicator function $R_j, j=1,\cdots,15$:

\[
\begin{array}{l}
R_1 = \frac{ 1 }{ 8 } (1 - X_{4 }) (1+ X_{ 1 2 }) ( 1 - X_{ 2 3 5 } ) \\
R_2 = \frac{ 1 }{ 8 } (1 + X_{1 }) (1+ X_{ 2 3 }) ( 1 + X_{ 2 4 5 } ) \\
R_3 = \frac{ 1 }{ 8 } (1 + X_{1 }) (1- X_{ 4 5 }) ( 1 - X_{ 2 3 5 } ) \\
R_4 = \frac{ 1 }{ 8 } (1 - X_{2 }) (1+ X_{ 3 4 }) ( 1 - X_{ 1 4 5 } ) \\
R_5 = \frac{ 1 }{ 8 } (1 + X_{2 }) (1+ X_{ 1 5 }) ( 1 - X_{ 3 4 5 } ) \\
R_6 = \frac{ 1 }{ 8 } (1 - X_{2 3 }) (1- X_{ 4 5 }) ( 1 - X_{ 1 3 5 } ) \\
R_7 = \frac{ 1 }{ 8 } (1 - X_{3 }) (1+ X_{ 2 5 }) ( 1 - X_{ 1 4 5 } ) \\
R_8 = \frac{ 1 }{ 8 } (1 + X_{3 }) (1+ X_{ 1 4 }) ( 1 - X_{ 2 4 5 } ) \\
R_9 = \frac{ 1 }{ 8 } (1 - X_{ 1 4 }) (1- X_{ 2 5 }) ( 1 + X_{ 3 4 5 } ) \\
R_{10} = \frac{ 1 }{ 8 } (1 - X_{1 5 }) (1- X_{ 3 4 }) ( 1 + X_{ 2 4 5 } ) \\
R_{11} = \frac{ 1 }{ 8 } (1 - X_{5 }) (1+ X_{ 1 3 }) ( 1 - X_{ 2 3 4 } ) \\
R_{12} = \frac{ 1 }{ 8 } (1 + X_{4 }) (1+ X_{ 3 5 }) ( 1 - X_{ 1 2 5 } ) \\
R_{13} = \frac{ 1 }{ 8 } (1 + X_{5 }) (1+ X_{ 2 4 }) ( 1 - X_{ 1 3 4 } ) \\
R_{14} = \frac{ 1 }{ 8 } (1 - X_{1 2 }) (1- X_{ 3 5  }) ( 1 + X_{ 2 4 5 } ) \\
R_{15} = \frac{ 1 }{ 8 } (1 - X_{1 3 }) (1- X_{ 2 4}) ( 1 + X_{ 3 4 5 } ) 

\end{array}
\]
To build a generic decomposition of $\fraction$ we start from one of these indicator function, let's say $R_1$ that identify the regular fraction $\mathcal R_1$. We have now to choose another indicator functions in the set made up by $R_2 , \dots, R_{ 1 5 }$, let's say $R_k$, with the condition that the corresponding regular fraction $\mathcal R_k$ doesn't intersect $\mathcal R_1$: $\mathcal R_1 \cap \mathcal R_k = \emptyset$. We have two possible choices, $R_{ 12 }$ and $R_{ 14 }$. If we choose $R_{ 12 }$ the only possible remaining is $R_{ 14 }$ and, viceversa, if we choose $R_{ 14 }$ the only possible remaining is $R_{ 12 }$.
Repeating the same procedure for all the $R_i$ and considering only the different decompositions, we get that $\fraction$ can be considered as the union of three regular 4-points designs
\[
\begin{array}{l}
\fraction = \mathcal R_1 \cup \mathcal R_{ 12 } \cup \mathcal R_{ 14 }  \\
\fraction = \mathcal R_2 \cup \mathcal R_3 \cup \mathcal R_6  \\
\fraction = \mathcal R_4 \cup \mathcal R_5 \cup \mathcal R_{ 10 }  \\
\fraction = \mathcal R_7 \cup \mathcal R_8 \cup \mathcal R_9  \\
\fraction = \mathcal R_{ 11 } \cup \mathcal R_{ 13 } \cup \mathcal R_{ 15 }  

\end{array}
\]

The decomposition that has been found in the previous section is $\fraction = \mathcal R_1 \cup \mathcal R_{ 12 } \cup \mathcal R_{ 14 }$.

\subsection{Decomposition of \emph{all} the ``Plackett-Burman'' designs with m=5 and 12 different runs into \emph{all} the unions of 4-points regular designs }
Using an ad-hoc software routine written in SAS IML we consider all the $\left( \begin{array}{c} 11 \\ 5 \end{array} \right) = 462$ different designs that can be obtained choosing 5 columns out of the 11 of the original designs. We get the following table where the first column contains an identification of the design, the second column the number of designs that are equal to the design and the third column the number of different runs contained in the design. For example, the design $\fraction$ that we have considered in the previous sections, belongs to the class ``69''. There are 11 designs that are equal to $\fraction$ and each has 12 points.

{\tiny 
\[
\begin{array}{|c|c|c|}
\begin{array}{rrr}
 ID & N & SIZE \\ \hline
     1 &  8 & 12 \\
     2 &  7 & 12 \\
     3 &  6 & 12 \\
     4 &  8 & 12 \\
     5 &  5 & 12 \\
     6 &  7 & 11 \\
     7 &  2 & 12 \\
     8 & 13 & 12 \\
     9 &  6 & 12 \\
    10 & 11 & 11 \\
    11 &  7 & 12 \\
    12 &  7 & 12 \\
    13 &  5 & 12 \\
    14 &  7 & 11 \\
    15 & 10 & 12 \\
    16 &  6 & 12 \\
    17 &  7 & 12 \\
    18 &  3 & 12 \\
    19 &  7 & 12 \\
    20 & 11 & 12 \\
    21 &  5 & 12 \\
    22 &  8 & 12 \\
    23 &  4 & 12 \\
    24 &  7 & 12 \\
    25 &  2 & 12 \\
    26 &  5 & 12 \\
    27 &  6 & 11 
    \end{array} &
		\begin{array}{rrr}
 		ID & N & SIZE \\ \hline
    28 &  6 & 12 \\
    30 & 10 & 12 \\
    32 &  6 & 11 \\
    35 &  6 & 12 \\
    37 &  3 & 12 \\
    39 &  4 & 12 \\
    44 & 11 & 12 \\
    45 &  7 & 12 \\
    46 &  6 & 12 \\
    49 &  2 & 12 \\
    51 &  7 & 12 \\
    52 &  9 & 12 \\
    53 &  5 & 12 \\
    54 &  4 & 11 \\
    55 &  4 & 12 \\
    57 &  3 & 11 \\
    58 &  6 & 12 \\
    61 &  6 & 12 \\
    63 &  4 & 12 \\
    64 &  3 & 12 \\
    65 &  8 & 12 \\
    66 &  5 & 12 \\
    67 &  2 & 12 \\
    68 &  7 & 12 \\
    69 & 11 & 12 \\
    70 & 13 & 12 \\
    71 &  6 & 12 
    \end{array} &
		\begin{array}{rrr}
		ID & N & SIZE \\  \hline
		72 &  5 & 11 \\
    73 &  6 & 12 \\ 
    74 &  5 & 12 \\
    82 &  6 & 12 \\
    84 &  2 & 12 \\
    85 &  9 & 11 \\
    87 &  7 & 12 \\
    89 &  4 & 12 \\
    94 &  6 & 12 \\
    98 &  7 & 12 \\
   100 &  3 & 12 \\
   101 &  8 & 12 \\
   102 &  3 & 11 \\
   103 &  7 & 12 \\
   110 &  2 & 12 \\
   116 &  5 & 12 \\
   117 &  1 & 12 \\
   128 &  2 & 12 \\
   134 &  5 & 12 \\
   140 &  3 & 12 \\
   146 &  5 & 11 \\
   147 &  3 & 12 \\
   149 &  4 & 12 \\
   154 &  6 & 12 \\
   159 &  1 & 12 \\
   167 &  2 & 12 \\
   184 &  1 & 12 
\end{array}
\end{array}
\]
}
It follows that the 462 designs can be partitioned into 81 classes:
\begin{itemize}
\item there are 70 classes where each design contains 12 runs
\item there are 11 classes where each design contains 11 runs
\end{itemize}

We limit to designs with 12 different runs.  We repeat the procedure described in the previous section for all the 70 different designs. First of all we determine the indicator functions of all the 70 designs. Every indicator function has the following form:
{
\begin{multline*}
\frac{ 3 }{ 8 } + a_{ 3 4 5 } X_{ 3 4 5 } + a_{ 2 4 5 } X_{ 2 4 5 } + a_{ 2 3 5 } X_{ 2 3 5 } +
 a_{ 2 3 4 } X_{ 2 3 4 } + \\ 
a_{ 2 3 4 5 } X_{ 2 3 4 5 } + a_{ 1 4 5 } X_{ 1 4 5 } + a_{ 1 3 5 } X_{ 1 3 5 } + a_{ 1 3 4} X_{ 1 3 4 } + 
a_{ 1 3 4 5 }  X_{ 1 3 4 5 } + a_{ 1 2 5 } X_{ 1 2 5 } + \\
a_{ 1 2 4 } X_{ 1 2 4 } + a_{ 1 2 4 5 } X_{ 1 2 4 5 } + a_{ 1 2 3 } X_{ 1 2 3 } + 
a_{ 1 2 3 5 }X_{ 1 2 3 5 } + a_{ 1 2 3 4} X_{ 1 2 3 4 }
\end{multline*}
}
where the coefficients $a_{ 3 4 5 }, \dots, a_{ 1 2 3 4}$ are equal to $\pm \frac { 1}{ 8}$.

We decompose every fraction into three 4-points regular design. 

As for the design considered in the previous example we have that every design
\begin{itemize}
\item contains $15$ ``4-points regular design'' 
\item can be considered as the union of three regular designs in 5 different ways
\end{itemize}

We have examined the decomposition structure of all the 70 designs. If we indicate with $R_1$, $R_2$ and $R_3$ the indicator functions of the regular designs contained into one of the design, we get

{\tiny 
\[
\begin{array}{lllll}
R_1 = \frac{1}{8} ( 1 & + e_1 X_{\alpha_1} & + e_2 X_{\alpha_2} & & + 
e_1 e_2 X_{\alpha_1 + \alpha_2} + e_4 X_{\alpha_4} + e_1 e_4 X_{\alpha_1 + \alpha_4} \\
& & & & + e_2 e_4 X_{\alpha_2 + \alpha_4} + e_1 e_2 e_4 X_{\alpha_1 + \alpha_2 + \alpha_4}  )
 \\

R_2 = \frac{1}{8} ( 1 & - e_1 X_{\alpha_1} & & + e_3 X_{\alpha_3} & - e_1 e_3 X_{\alpha_1 + \alpha_3} + e_5 X_{\alpha_5} - e_1 e_5 X_{\alpha_1 + \alpha_5} \\
& & & & + e_2 e_5 X_{\alpha_2 + \alpha_5} + e_1 e_3 e_5 X_{\alpha_1 + \alpha_3 + \alpha_5}  )\\
R_3 = \frac{1}{8} ( 1 & & - e_2 X_{\alpha_2} & - e_3 X_{\alpha_3} & + e_2 e_3 X_{\alpha_2 + \alpha_3} + e_6 X_{\alpha_6} - e_2 e_6 X_{\alpha_2 + \alpha_6} \\
& & & & - e_3 e_6 X_{\alpha_3 + \alpha_6} + e_2 e_3 e_6 X_{\alpha_2 + \alpha_3 + \alpha_6} )

\end{array}
\]
} 
where,  being  $|\alpha| = \sum_j \alpha_j$,   

\begin{itemize}
  \item $ |\alpha_1|$, $| \alpha_2 |$ and $| \alpha_3 |$  are all less than three
  \item all the others, i.e. $| \alpha_1 +  \alpha_2  |$ , $\dots$, $| \alpha_2 + \alpha_3 + \alpha_6 |$ are all greater or equal to 3
\end{itemize}
  
This evidence has suggested the following procedure.
  \begin{enumerate}
  \item We have built all the $\alpha_1, \dots, \alpha_6$ that satisfy the previous requirement,
  \[
\begin{array}{l|cccccc}
N & \alpha_1 & \alpha_2 & \alpha_3 & \alpha_4 & \alpha_5 & \alpha_6 \\
1 &    1 & 2 3 & 4 5 & 2 4 5 & 2 3 4 & 1 2 4 \\
2 &		1 & 2 4 & 3 5 & 2 3 5 & 2 3 4 & 1 2 3 \\
3 &    1 & 2 5 & 3 4 & 2 3 4 & 2 3 5 & 1 2 3 \\
4 &    2 & 1 3 & 4 5 & 1 4 5 & 1 3 4 & 1 2 4 \\
5 &    2 & 1 4 & 3 5 & 1 3 5 & 1 3 4 & 1 2 3 \\
6 &    2 & 1 5 & 3 4 & 1 3 4 & 1 3 5 & 1 2 3 \\
7 &    3 & 1 2 & 4 5 & 1 4 5 & 1 2 4 & 1 3 4 \\
8 &    3 & 1 4 & 2 5 & 1 2 5 & 1 2 4 & 1 2 3 \\
9 &    3 & 1 5 & 2 4 & 1 2 4 & 1 2 5 & 1 2 3 \\
10 &    4 & 1 2 & 3 5 & 1 3 5 & 1 2 3 & 1 3 4 \\
11 &    4 & 1 3 & 2 5 & 1 2 5 & 1 2 3 & 1 2 4 \\
12 &    4 & 1 5 & 2 3 & 1 2 3 & 1 2 5 & 1 2 4 \\
13 &    5 & 1 2 & 3 4 & 1 3 4 & 1 2 3 & 1 3 5 \\
14 &    5 & 1 3 & 2 4 & 1 2 4 & 1 2 3 & 1 2 5 \\
15 &    5 & 1 4 & 2 3 & 1 2 3 & 1 2 4 & 1 2 5
\end{array}
\]
\item For every choice of $\alpha_1, \dots, \alpha_6$ we have built the 64 indicator functions that correspond to all the values of $e_1, \dots, e_6$, being $e_i = \pm 1, i=1,\dots, 6$.
\end{enumerate}  

According to this procedure we have generated $15 \times 64 =960$ indicator functions. If we limit to the different ones we get \emph{192 indicator functions}. This number is the same that has been found in \cite{carlini|pistone:2007}, as the total number of orthogonal arrays of strength 2.

\subsection{Remark}
It is interesting to point out that the ``understanding'' of the mechanism underlying the Plackett-Burman designs (m=5, 12-runs) has allowed to build \emph{all } the orthogonal arrays of strength 2.

%\begin{subsection}{Latin Squares}
%Another example can be considered in the class of latin squares. \marginpar{ serve? }
%\end{subsection}

\section{Conclusions}
  \begin{itemize}
\item The problem to determine regular designs that are contained in a given fraction has been faced.
\item A condition in terms of the coefficients of the polynomial indicator function has been found.
\item The decomposition of a given fraction into regular designs seems useful for fractional factorial generation.
\end{itemize}  
\bibliographystyle{plain}
%\bibliography{/Users/gianni/Archive/bibs/tutto}
%\bibliography{/home/pistone/Archive/bibs/tutto}
%\bibliography{C:/tutto}
%\bibliography{C:/tutto1}

\end{document}